\documentclass[12pt]{article}
\usepackage{a4}
\usepackage{amsfonts}
\usepackage{amssymb}
\usepackage{epsfig}

\newcommand{\be}{\begin{equation}}
\newcommand{\ee}{\end{equation}}
\newcommand{\bea}{\begin{eqnarray}}
\newcommand{\eea}{\end{eqnarray}}
\newcommand{\mbb}{\mathbb}
\newcommand{\ti}{\times}
\newcommand{\half}{\frac{1}{2}}

\newcommand{\mc}{\mathcal}

\begin{document}

\title{
\begin{flushright} \vspace{-2cm}
{\small DAMTP-2006-17 \\ \vspace{-0.35cm}
hep-th/0602233} \end{flushright}
\vspace{3cm}
{\bf The QCD Axion and Moduli Stabilisation}
}
\author{}
\date{}

\maketitle

\begin{center}
Joseph P. Conlon\footnote{email:j.p.conlon@damtp.cam.ac.uk}

\bigskip\medskip
DAMTP, Centre for Mathematical Sciences, Wilberforce Road, Cambridge, CB3 0WA, UK.

\end{center}

\medskip

\begin{abstract}
\noindent

We investigate the conditions for a QCD axion to coexist with
stabilised moduli in string compactifications.
We show how the simplest approaches to moduli stabilisation give
unacceptably large masses to the axions. We observe that solving the F-term
equations is insufficient for realistic moduli stabilisation and give a no-go
theorem on supersymmetric moduli stabilisation with unfixed axions
applicable to all string compactifications and relevant to much current work.
We demonstrate how nonsupersymmetric moduli stabilisation with unfixed
axions can
be realised. We finally outline how to stabilise the moduli such that
$f_a$ is within the allowed window $10^9 \textrm{GeV} < f_a < 10^{12}
\textrm{GeV}$, with $f_a \sim \sqrt{M_{SUSY} M_P}$.

\end{abstract}

\thispagestyle{empty}
\clearpage

\tableofcontents

\section{Introduction}
\linespread{1.2}

String theory remains certainly the most popular and on all appearances the best candidate for an ultraviolet
completion of the Standard Model that will unify gauge and gravitational interactions in a consistent
quantum theory. Since the discovery of Calabi-Yau compactifications
\cite{chsw},
it has been hoped that the measured properties of the Standard Model
may be understood through the details of a string compactification.

This has traditionally been approached top-down, starting with a
fully specified and globally consistent closed string compactification.
The defined technical problem is to construct a compactification
whose low energy gauge group and matter content is that of the Standard Model or a close extension.
The original context of this is the weakly coupled heterotic string on a Calabi-Yau, where the
matter spectrum is determined by the geometry of the manifold and the gauge bundles thereon.
Realistic matter content requires substantial mathematical effort;
for an account of some recent developments \cite{hepth0505041} can be consulted.

The style of top-down constructions changed with the discovery of D-branes and the realisation that
Standard-like models could also be constructed within `intersecting brane worlds'.
In these scenarios the Standard Model is located on brane stacks that wrap cycles in the internal space, with
the matter content determined by the intersection numbers of these cycles.
Thus it is not just heterotic, but also (orientifolded) type II compactifications that can give
pseudo-realistic physics.
Such constructions have been
developed both for toroidal orbifolds and smooth Calabi-Yaus, with a recent review being \cite{hepth0502005}.

A general problem in top-down constructions is that of moduli stabilisation. The simplest string compactifications
have very many moduli - massless scalar fields coupling with gravitational strength and determining the matter coupling constants.
Such massless particles are however inconsistent with experiment and it is necessary to generate a potential for them. The
difficulty is that moduli are associated with the geometry of the compactification, and the effects - preeminently fluxes -
that stabilise the moduli also back-react on the geometry, making the resulting space difficult to study.

Together with the brane world picture, this has led to `bottom-up'
approaches to string
phenomenology \cite{hepth0005067}. As branes
are localised, the field theory on them depends only on local geometry.
A bottom-up physicist first builds a local Standard Model, and only later worries about the
global embedding.
The limiting case of this is `branes at singularities', where the entire low energy spectrum is determined by the nature
of a pointlike singularity. In this context
there have been recent attempts to find a Standard Model singularity \cite{hepth0508089}.
A characteristic of this approach is that particle physics becomes an open string theory decoupled from the
closed string dynamics associated with the compact geometry.

Regardless of the philosophy, any realistic string compactification must eventually
account for the structure of the Standard Model.
In this respect moduli stabilisation connects top-down and bottom-up constructions, as the moduli vevs determine the
field theory coupling constants.
One Standard Model coupling constant in particular need of explanation is the QCD $\theta$ angle.
This is of course a well-known problem with a well-known answer: a Peccei-Quinn axion
\cite{PecceiQuinn}.
There do exist other possibilities, reviewed in \cite{hepph0011376}, but for this paper I shall simply assume
the Peccei-Quinn solution to be correct.

In the context of string theory, this creates a modulus anti-stabilisation problem.
There are many string theory axions that may in principle solve the strong CP problem.
To do so, an axion must remain massless
throughout the thicket of moduli stabilisation effects and down to the
QCD scale. This problem has previously been brought up in \cite{hepth0309170,hepth0310203}.
It is clean and sharply posed, as effects very weak
on the Planck scale may be very large on the QCD scale.
Given the necessity of moduli stabilisation, this question is best analysed within the context of
string constructions for which all moduli have been stabilised. The purpose of this paper is to
determine for such constructions the conditions under which a QCD axion will exist.

The paper is organised as follows. I first review in section \ref{StrongCP} the strong CP problem and the ways axions can arise
in different string theory constructions. Section
\ref{ModuliStabilisation} investigates how to
stabilise moduli while keeping axions sufficiently light to solve the
strong CP problem.
I review some approaches to stabilising all moduli
and show why their simplest versions contain no light axions.
I investigate potentials with massless axions and show that
in such cases there exist no supersymmetric minima of the potential,
but nonsupersymmetric minima may exist.
I give conditions on the geometry for a QCD axion to exist, both at
leading and higher orders in the instanton expansion.
Section \ref{AxionDecayConstantSec} addresses the axion decay
constant. In the context of the exponentially large
volume compactifications of \cite{hepth0502058, hepth0505076} I
outline how a phenomenologically acceptable value of $f_a$ may be
achieved together with the relationship $f_a \sim \sqrt{M_{SUSY} M_P}$.

\section{Axions}
\label{StrongCP}

\subsection{Axions and the Strong CP Problem}

In principle, the strong interactions can generate CP violation
through an $F \tilde{F}$ coupling,
\be
S_{F\tilde{F}} = \frac{\theta}{16 \pi^2} \int F_{\mu \nu} \tilde{F}^{\mu \nu}.
\ee
However, observationally $\theta$ is extremely small: $\vert \theta \vert < 10^{-8}$. Given that at the above level,
$\theta$ is a coupling constant which can \emph{a priori} take values anywhere between 0 and $2 \pi$, this
seems unnatural. The strong CP problem is to explain this observation.

There exist several proposed resolutions. In the context of string theory and string compactifications, the most natural
is that due to Peccei and Quinn \cite{PecceiQuinn}. In this approach $\theta$ is promoted to a dynamical field, with Lagrangian
\be
\label{axionLagrangian1}
\mc{L} = \mc{L}_{SM} + \frac{1}{2} f_a^2 \partial_\mu \theta \partial^\mu \theta + \frac{\theta}{16 \pi^2} F_{\mu \nu} \tilde{F}^{\mu \nu}.
\ee
In (\ref{axionLagrangian1}) $f_a$ has dimensions of mass and is known as the axion decay constant.
Conventionally a scalar has mass dimension one,
and so we redefine
$a \equiv \theta f_a$. This gives
\be
\label{axionLagrangian}
\mc{L} = \mc{L}_{SM} + \frac{1}{2} \partial_\mu a \partial^\mu a + \frac{a}{16 \pi^2 f_a} F_{\mu \nu} \tilde{F}^{\mu \nu}.
\ee
In equation (\ref{axionLagrangian}) there exists an anomalous global $U(1)$ symmetry, $a \to a + \epsilon$.
This symmetry is violated by QCD instanton effects, which break it to a discrete subgroup.
These generate a potential for $a$,
\be
V_{\textrm{instanton}} \sim \Lambda_{QCD}^4 \left( 1- \cos \left( \frac{a}{f_a} \right) \right).
\ee
In the absence of other effects, this potential is minimised at $a=0$, setting the $\theta$-angle to zero dynamically.
The mass scale for the axion $a$ is
\be
\label{axionMasses}
m_a \sim \frac{\Lambda_{QCD}^2}{f_a}.
\ee
(A more precise estimate replaces $\Lambda_{QCD}^2$ by $f_{\pi} m_{\pi}$, where $f_{\pi} \sim 90 \textrm{MeV}$ is the pion
decay constant.)
The Peccei-Quinn symmetry is by its nature anomalous. For the axion to solve the strong CP problem, the leading anomalous
contribution to the potential must be that of QCD instantons,
otherwise the minimum will be at $\theta \neq 0$.

Phenomenologically, $f_a$ has only a narrow window of allowed values. The smaller the value of $f_a$, the more strongly the axion
couples to matter. The condition that supernovae cool by predominantly emitting
energy through neutrinos (rather than axions) is equivalent to the constraint
$f_a > 10^9 \textrm{GeV}$. There is also a cosmological upper bound on $f_a$.
The axion field presumably starts its cosmological evolution with $\theta$ at some arbitrary value between 0 and $2 \pi$.
Once the Hubble scale is comparable to the axion mass, the axion field oscillates and the energy density
stored in the axion field is diluted with the expansion of the universe.
The energy stored today thus depends on the axion mass, which is determined by $f_a$. The requirement that axions do not
overclose the universe leads to the constraint
$f_a < 10^{12} \textrm{GeV}$, with such axions being potential dark
matter candidates.

The requirement $f_a < 10^{12} \textrm{GeV}$ follows from the
assumption of a standard cosmology. String compactifications generally
have many moduli, which in supersymmetric scenarios typically have
masses $m \sim 1 \hbox{TeV}$. These can be long-lived, causing
problems with nucleosynthesis and giving low reheat temperatures. In
supersymmetric models, the axion always has a scalar partner (the
saxion) with the associated cosmological problems. It has been argued
(in particular see \cite{hepth9608197, hepph0210256}) that the
cosmological problems with the scalars are more severe than those
associated with the axion, and so the upper bound on $f_a$ should not
be taken seriously without an associated resolution of the
cosmological moduli problems. \cite{hepph0210256} reports that models
can be found in which
both axion and saxion cosmological problems may be evaded with
$f_a \sim 10^{15} \hbox{GeV}$.

The cosmological history of the universe before nucleosynthesis is not
known. However we are going to assume a standard cosmology and take
the upper bound on $f_a$ seriously. This is mainly because
low reheat temperatures ($T \sim 10 \hbox{MeV}$) have a generic problem that at reheating the
decaying modulus has an $\mc{O}(1)$ branching fraction to gauginos and
thus overproduces LSPs \cite{hepph0602061, hepph0602081}. In addition,
low reheat temperatures make it hard to produce the observed baryon asymmetry.
This suggests the universe should be hot at the weak scale, allowing a standard
WIMP annihilation calculation and the possibility of electroweak baryogenesis, but also imposing the standard bounds on
$f_a$. Furthermore, this is not in obvious contradiction with the properties of the saxion -
its matter coupling is also set by $f_a$ (rather than $M_P$), and low values of $10^9
\hbox{GeV} < f_a < 10^{12} \hbox{GeV}$ can easily give
large reheating temperatures $T \sim 10^5 \hbox{GeV}$.

\subsection{Axions in String Theory}

String compactifications generically contain fields $a_i$ which have
$a_i F \tilde{F}$ couplings and
possess the anomalous global symmetry $a \to a + \epsilon$ featuring
in the axionic solution to the strong CP problem.
We first enumerate possible axions, before considering their relation to the physics of moduli
stabilisation.

\subsubsection*{Axions in the Heterotic String}

In heterotic compactifications, the axions are traditionally divided into the universal, or model-independent,
axion and the model-dependent axions. The model-independent axion is the imaginary part of the dilaton multiplet, $S =
e^{-2 \phi} \mc{V} + i a$. It is the dual of the 2-form potential $B_{2,\mu \nu}$ arising from the NS-NS 2-form
field: $da = e^{-2 \phi} *dB_{\mu \nu}$. The dilaton superfield is the tree-level gauge kinetic function for
all gauge groups:
$$
\mc{L} \sim \textrm{Re}(S) \int F^a_{\mu \nu} F^{a \mu \nu} + \textrm{Im}(S) \int F^a_{\mu \nu} \tilde{F}^{a \mu \nu}.
$$
Consequently in a realistic compactification there must always exist an $a F_{QCD} \tilde{F}_{QCD}$ coupling.

There are also the model independent axions, $b_i$, given by the
imaginary parts of the K\"ahler moduli $T_i$.
For a basis $\Sigma_i$ of 2-cycles of the Calabi-Yau,
$T_i = t_i + i b_i$, with $t_i = \int_{\Sigma_i} J$ the string frame
volume of the cycle $\Sigma_i$ and
$b_i = \int_{\Sigma_i} B_2$.
At tree level, these have no couplings to QCD.
However, such a coupling may be generated through
 the one loop correction to the gauge kinetic function.
For the $E_8 \ti E_8$ heterotic string, this correction is
\bea
\label{HetOneLoop}
f_1 & = & S + \beta_i T_i, \\
f_2 & = & S - \beta_i T_i,
\eea
where $1$ and $2$ refer to the first and second $E_8$ respectively.
The $\beta_i$ are determined by
the gauge bundles on the compactification manifold $X$.
For gauge bundles $V_1$ and $V_2$,
\be
\beta_i = \frac{1}{8 \pi^2} \int e_i \wedge  \left( c_2(V_1) -
c_2(V_2) \right),
\ee
where $e_i$ is the 2-form associated with the cycle $\Sigma_i$.
This can be derived by dimensional reduction of the Green-Schwarz term
$\int B_2 \wedge X_8(F_1, F_2, \mc{R})$.\footnote{Strictly, this only gives the correction to $\textrm{Im}(f)$. The
corresponding correction to $\textrm{Re}(f)$ is however implied by holomorphy.} The axions associated to the K\"ahler moduli
are called model-dependent as their appearance in $f$ depends on the one-loop correction, which in turn depends on the specific
properties of the compactification.

\subsubsection*{Axions in Intersecting Brane Worlds}

The discovery of D-branes \cite{Polchinski} led to the extension of
string model building beyond the heterotic string. The type II string theories,
or more properly orientifolds thereof, can give rise to `intersecting brane worlds'. In these the
standard model is localised on a stack of branes while gravity propagates in the bulk. Light bifundamental matter arises
from strings located at intersection loci and stretching between brane stacks.

The bosonic action of a single Dp-brane is
\be
S_p =  \frac{- 2 \pi}{(2 \pi \sqrt{\alpha'})^{p+1}} \left(
\int_{\Sigma} d^{p+1} \xi e^{-\phi} \sqrt{ \textrm{det}(g + B + 2 \pi \alpha' F)}
+ i \int_{\Sigma} e^{B + 2 \pi \alpha' F} \wedge \sum_q C_q \right).
\ee
$\Sigma$ is the cycle wrapped by the brane and the sum is a formal sum
over all
RR potentials in which only relevant terms are picked out.

The kinetic term $F_{\mu \nu} F^{\mu \nu}$ comes from the DBI action
and the instanton action $F \wedge F$ from the Chern-Simons term.
The gauge coupling corresponds to the inverse volume of $\Sigma$ and the $\theta$ angle to the
component of $C_{p-3}$ along $\Sigma$. These fields pair up to become the scalar component of the chiral multiplet which is
the gauge kinetic function of the resulting gauge theory.

As IIB compactifications are our main focus, we shall be more explicit here. In principle, IIB string theory
allows, consistent with supersymmetry, space-filling D3, D5, D7 and D9-branes. However, in an orientifold setting we are restricted
to either D3/D7 or D5/D9 pairs. We shall interest ourselves in the former case. The branes must wrap appropriate cycles to cancel
the negative charge and tension carried by the orientifold planes; we assume this has been done.

In D3/D7 compactifications, the relevant superfields
are those of the dilaton and K\"ahler moduli multiplets. Their scalar
components are\footnote{Technically,
this is only for the case that $h^{1,1}_{-} = 0$. This will not be
important for
the issues we discuss, and so we will use this
simplifying assumption. The correct expressions under more general
circumstances can
be found in \cite{hepth0403067, hepth0502059}.}
\bea
S & = & e^{-\phi} + i c_0, \\
T_i & = & \tau_i + i c_i.
\eea
$c_0$ is the Ramond-Ramond 0-form and $e^\phi \equiv g_s$ the string coupling. For $\Sigma_i$ a 4-cycle of the Calabi-Yau,
$$
c_i = \frac{1}{l_s^4} \int_{\Sigma_i} C_4 \qquad \textrm{ and } \qquad
\tau_i = \int_{\Sigma_i} \frac{e^{-\phi}}{2} J \wedge J,
$$
where $l_s = 2 \pi \sqrt{\alpha'}$ denotes the string length.
Note that the K\"ahler modulus $T_i$ involves the Einstein, rather
than string, frame volume of a 4-cycle.
This is most simply
understood as the requirement that the gauge kinetic function be holomorphic in the chiral superfields.
Indeed, $S$ is the universal gauge kinetic function for D3-branes,
whereas $T_i$ is the gauge kinetic function for the field theory on a D7-brane stack
wrapping the 4-cycle $\Sigma_i$.

The axionic couplings arise from the Chern-Simons term in the action. For D3-branes, this gives
\be
S_{F \tilde{F}} = \frac{c_0}{4 \pi} \int F \wedge F,
\ee
while for D7-branes
\be
S_
{F \tilde{F}} = \frac{c_i}{4 \pi} \int F \wedge F.
\ee
By expanding the DBI action, we obtain the field theory couplings
$$
\frac{1}{g^2}\Bigg\vert_{D3} = \frac{e^{-\phi}}{2 \pi} \qquad \textrm{
  and }
\qquad \frac{1}{g^2}\Bigg\vert_{D7} = \frac{\tau_i}{2 \pi}.
$$

There exists a similar story for IIA intersecting brane worlds, where the Standard Model must be realised
on wrapped D6-branes (a Calabi-Yau has no 1- or 5-cycles to wrap D4- or D8-branes on).
The gauge coupling now comes from the calibration form $\textrm{Re}(\Omega)$ and
the axion from the reduction of the 3-form potential $C_3$,
$$
\frac{1}{g^2}\Bigg\vert_{D6} = \int_{\Sigma_i} \textrm{Re}(\Omega) \qquad \textrm{ and } \qquad \theta \Bigg\vert_{D6} =
\int_{\Sigma_i} C_3.
$$
Our interest is in the interaction of axions with moduli stabilisation
and supersymmetry breaking, to which we now turn.

\section{Axions and Moduli Stabilisation}
\label{ModuliStabilisation}

It is obvious from the above that potential
axions are easily found in string compactifications; indeed, they are superabundant.
For axions to solve the strong CP problem, they must also be light, with QCD instantons giving the dominant
contribution to their potential.
Light axions are not in themselves
problematic. Pure type II Calabi-Yau compactifications have
many axions, which remain exactly massless as a consequence of four-dimensional $\mc{N}=2$ supersymmetry.
However, the same $\mc{N}=2$ supersymmetry that guarantees the axions remain massless also
guarantees a non-chiral spectrum with the axions' scalar partners massless.
These are modes of the
graviton and will lead to unobserved fifth forces.

More realistic string constructions have $\mc{N}=1$ supersymmetry in four dimensions,
allowing a potential to be generated for the moduli.
To avoid bounds from fifth force experiments, the size moduli - the saxions - must at a minimum receive masses at a scale
$m_T \gtrsim (100 \mu \textrm{m})^{-1}$. However, the expected scale is much larger: typical constructions
give moduli masses comparable to the supersymmetry breaking scale, and the
cosmological moduli problem \cite{hepph9308292, hepph9308325} suggests that in fact
$m_T \gtrsim 10 \textrm{TeV}$.
In recent years there has been much progress
in moduli stabilisation \cite{gkp, kklt}, with physics such as fluxes and instantons used to lift the degeneracies associated
with the geometric moduli.
In the context of strong CP, this same progress creates a modulus
anti-stabilisation problem.
There are many stringy effects that can generate
a potential for a putative QCD axion. These include worldsheet instantons, D-instantons and
gaugino condensation, all of which are often invoked to stabilise the geometric moduli present.
If any one of these effects is more important for a given axion than
QCD instantons, that axion does not solve the strong CP problem.

Axion potentials generally come from nonperturbative effects whose magnitude is exponentially sensitive
to the values of the stabilised moduli. The analysis of such effects is therefore best
performed in a context within which all moduli are stabilised, and we first review mechanisms to
achieve this. The most developed scenarios are those within IIB flux compactifications.
We will work mostly with these, although we will along the way obtain a no-go
theorem applicable to all string compactifications. IIB compactifications have the advantage that
the stabilisation of dilaton and complex structure moduli
can be studied ten-dimensionally \cite{gkp}, while the back-reaction of the fluxes is relatively mild:
the internal geometry simply becomes conformally Calabi-Yau.

\subsection{Review of Moduli Stabilisation}

We shall orient our subsequent discussion around two approaches to stabilising all moduli, the well-known
KKLT scenario \cite{kklt} and the exponentially large volume models
of \cite{hepth0502058, hepth0505076}.
The point of departure is the flux compactifications of Giddings, Kachru and Polchinski \cite{gkp}. These are compactifications of
a IIB orientifold\footnote{or more generally F-theory.} in the presence of
localised sources (D3/D7 branes and O3/O7 planes) with non-vanishing 3-form flux $G_3 = F_3 - S H_3$.
Here $F_3$ and $H_3$ are RR and NSNS 3-form fluxes with $S$ the dilaton-axion.
The fluxes generate a superpotential\cite{hepth9906070}
\be
W = \frac{1}{l_s^2} \int G_3 \wedge \Omega,
\ee
with $\Omega$ the unique holomorphic (3,0) form and the K\"ahler potential given by
\be
\frac{\mc{K}}{M_P^2} = - 2 \ln(\mc{V}) - \ln \left( \int \Omega \wedge \bar{\Omega} \right) - \ln (S + \bar{S}).
\ee
This K\"ahler potential is no-scale for the K\"ahler moduli.
The dilaton and complex structure moduli appear in the superpotential and are stabilised at
leading order in the $g_s$ and $\alpha'$ expansions. Their masses scale as
\be
\label{DilatonMass}
m_S \sim m_{\phi} \sim \frac{N}{R^6} M_P,
\ee
where $N$ is a measure of the number of units of 3-form flux and $R$
is the Calabi-Yau radius in units of $l_s$.

At this level, the no-scale property implies the K\"ahler moduli remain unstabilised.
The dilaton and complex structure moduli are integrated out to focus on an effective theory for the K\"ahler moduli.
Although absent at tree level, the K\"ahler moduli can appear nonperturbatively in the superpotential
through brane instantons \cite{hepth9604030} or gaugino condensation,
\be
W = W_0 + \sum_i A_i e^{-a_i T_i},
\ee
where $a_i$ = $2 \pi (\frac{2 \pi}{N})$ for brane instantons (gaugino condensation).
The KKLT proposal is to stabilise the K\"ahler moduli by solving $D_{T_i} W = 0$ for all $i$. The resulting 4-cycle sizes are
\be
\tau_i \sim \frac{1}{a_i} \ln \left( \frac{W_0}{A_i} \right).
\ee
The cycle size goes logarathimically with $W_0$, and so to trust the supergravity approximation
$W_0$ must be extremely small.
We may usefully rephrase this condition. The KKLT construction relies on nonperturbative corrections to the scalar
potential dominating the perturbative corrections. This is equivalent to the requirement that the `correction' to the
superpotential is comparable to, or larger, than the tree-level
term. In order
for this to hold at (moderately) large $T$, $W_0$ must be
very small, which is to be achieved by tuning fluxes.

However, for almost all values of $W_0$ (`almost' can be made precise as in \cite{hepth0404116})
 there exists no reliable regime in which this requirement is met. Perturbative corrections,
originating from corrections to the K\"ahler potential, are then essential for the study of the moduli potential. It was shown in
\cite{hepth0502058, hepth0505076}
 that the incorporation of an $\alpha'^3$ correction to the K\"ahler potential \cite{hepth0204254},
\be
\label{KahlerCorrection}
\mc{K} = - 2 \ln (\mc{V}) \to \mc{K} = -2 \ln \left(\mc{V} + \frac{\xi}{2 g_s^{\frac{3}{2}}}\right),
\ee
combined with the same nonperturbative superpotential corrections
$$
W = W_0 + \sum_{i=2}^{h^{1,1}} A_i e^{-a_i T_i},
$$
 leads, subject to a necessary condition $h^{2,1} > h^{1,1} > 1$,
 to a minimum of the potential at exponentially large volumes:
$$
\mc{V} \sim W_0 e^{\frac{c}{g_s}},
$$
for a model-dependent constant $c$. This and other properties of the minimum follow from
an explicit study of the scalar potential, but a summary of the results is as follows:
\begin{enumerate}
\item The moduli divide into one large modulus $\tau_b$ and $h^{1,1} -1 $ small `blowup' moduli $\tau_i$.
Whereas the former is exponentially large, $\tau_b \sim \mc{V}^{\frac{2}{3}}$,
the latter all have $\tau_i \sim \ln(\mc{V})$. The origin of this is as follows.
If we take $\tau_b^{\frac{3}{2}} \sim \mc{V} \gg 1$ and fixed, the other moduli $\tau_i$ minimise their potential at
$D_{T_i} W = 0 + \mc{O}\left(\frac{1}{\mc{V}}\right)$. Neglecting numerical factors, this
generates an effective potential for $\mc{V}$ of
$$
V \sim - \frac{(\ln \mc{V})^{\frac{3}{2}}}{\mc{V}^3} + \frac{\xi}{g_s^{\frac{3}{2}} \mc{V}^3},
$$
where the $\frac{\xi}{\mc{V}^3}$ comes from the K\"ahler correction. $\mc{V}$ is then dynamically
stabilised at $\mc{V} \gg 1$.
\item
The stabilised volume is exponentially sensitive to the stabilised string coupling, allowing a natural generation
of hierarchies: the structure of the potential is such as to create a hierarchical separation of the string and Planck scales.
\end{enumerate}

The K\"ahler correction (\ref{KahlerCorrection}) - of course only one of many - plays
an essential role in this construction. The effects of other possible
K\"ahler corrections, such as from
warping effects or other IIB
$\mc{O}(\alpha'^3)$ corrections, have been
extensively discussed in \cite{hepth0505076} (see also \cite{hepth0508171}).
We will comment further on these in the appendix.
Here we simply note that the only hope of incorporating K\"ahler corrections in a controlled manner
is via a small expansion parameter. In this case $\mc{V} \gg 1$ and
the expansion parameter is
$\frac{1}{\mc{V}}$. Of known corrections,
(\ref{KahlerCorrection}) gives the leading contribution to the scalar potential in the $\frac{1}{\mc{V}}$ expansion.

Finally let us note that the two constructions above
both give rise to AdS minima, which are respectively supersymmetric and non-supersymetric.
It is necessary to include a phenomenological uplift term, arising
from e.g.
a $\overline{\textrm{D3}}$-brane in a warped throat, in order
to match the observed cosmological constant. In the KKLT scenario, it is
the uplift that gives rise to soft supersymmetry breaking terms.
For the exponentially large volume compactifications, the
supersymmetry breaking is
inherited from the no-scale structure and the effects
of the uplift are subdominant \cite{hepth0505076, hepph0512081}.

\subsection{The Simplest Scenarios: Why Axions are Heavy}
\label{KKLTAxionSection}

We now turn to the axions.
Our first point is that the simplest versions of the above scenarios lack a QCD axion. All potential axions
receive a high scale mass and thus cannot solve the strong CP problem. For simplicity we concentrate on the
KKLT construction, but a very similar argument holds for the exponentially large volume compactifications.

We start by asking whether QCD is to be realised on D3 or D7 branes.
If we were to use D3-branes, the QCD axion would be
the imaginary component of the dilaton multiplet, $S = e^{-\phi} + i c_0$.
However, as indicated in (\ref{DilatonMass}) this multiplet is
stabilised at tree-level by the fluxes, with a mass $m_S \sim
\frac{N}{R^6} M_P$, with $R$ the radius in units of $l_s$.
This tree-level stabilisation may seem at odds with the axionic shift
symmetry $c_0 \to c_0 + 2 \pi$. However, the shift
symmetry is a subgroup of the fundamental $SL(2,Z)$ symmetry, under which the fluxes also transform.
By nature a duality, this is invisible in the low-energy theory and cannot protect the axion from acquiring a mass.
As $R \lesssim 5$ in KKLT, the axion obtains a mass $m_{c_0} \gtrsim 10^{15} \textrm{GeV}$ and cannot be a QCD axion.
For the exponentially large volume compactifications, $R$ is larger but the conclusion unchanged: QCD on a D3 brane stack is
inconsistent with the existence of a Peccei-Quinn axion.

This implies that QCD ought to be realised
on D7 branes. The axions are now the imaginary parts of the K\"ahler moduli,
and the instanton effects used to stabilise these moduli will also give the axions a mass.
To estimate the scale of this mass,
it is simplest just to construct the potential explicitly.

As above, we take the superpotential
\be
W = W_0 + \sum_{i=1}^{h^{1,1}} A_i \exp (-a_i T_i),
\ee
with K\"ahler potential
\be
\label{KahlerPotential}
\mc{K} = - 2 \log \mc{V}.
\ee
Note that $\mc{K} = \mc{K}(T_i + \bar{T}_i)$, and so the axions do not
appear either in $\mc{K}$ or its derivatives.
The supergravity F-term potential is
\be
\label{FTermPot}
V = e^{\mc{K}} (\mc{K}^{i \bar{j}} D_i W D_{\bar{j}} \bar{W} - 3 \vert W \vert^2).
\ee
The no-scale property of the K\"ahler potential simplifies (\ref{FTermPot}) to
\be
\label{NoScaleScalarPotential}
V = e^{\mc{K}} \left( \mc{K}^{i \bar{j}} \partial_i W
\partial_{\bar{j}} \bar{W} +
\mc{K}^{i \bar{j}}\left( (\partial_i \mc{K}) W \partial_{\bar{j}}
\bar{W} + (\partial_{\bar{i}} \mc{K} ) \bar{W} \partial_j W \right) \right).
\ee
It is a property of the K\"ahler potential (\ref{KahlerPotential}) that $\mc{K}^{i \bar{j}} \partial_i \mc{K} = -2 \tau_j$.
(\ref{NoScaleScalarPotential}) becomes
\bea
V & = & e^{\mc{K}} \Bigg( \mc{K}^{i \bar{j}} a_i a_j \left( A_i
\bar{A}_j e^{-a_i T_i -a_j \bar{T}_j} + \bar{A_i} A_j e^{-a_i \bar{T}_i -a_j \bar{T}_j} \right)
  \nonumber  \\
  & & - 2 a_i \tau_i
\left( W \bar{A}_i e^{-a_i \bar{T}_i} + \bar{W} A_i e^{-a_i T_i} \right) \Bigg) .
\label{LatestPotential}
\eea
It is easy to extract the axionic dependence of the potential (\ref{LatestPotential}).
$\mc{K}$ and its derivatives are all real and phases only come
from the superpotential. The potential becomes
\bea
V & = & e^{\mc{K}} \left( \mc{K}^{i \bar{j}} \left( 2 a_i a_j \vert
A_i A_j \vert e^{-a_i \tau_i - a_j \tau_j}
\cos(a_i \theta_i + a_j \theta_j + \gamma_{ij}) \right) \right. \\
& & - 4 a_i \tau_i \vert W_0 A_i \vert e^{-a_i \tau_i} \cos(a_i \theta_i + \beta_i) - 4 a_i \tau_i \vert A_i A_j \vert \cos (a_i \theta_i + a_j \theta_j +
\gamma_{ij}) \Big). \nonumber
\eea
$\theta_i$ denote the axions and the
phases $\gamma_{ij}$ and $\beta_i$ come from the phases of $A_i \bar{A}_j$ and $\bar{A}_i W$ respectively.
The axionic mass matrix is
\be
M^2_{ij} = \frac{\partial^2 V}{\partial \theta_i \partial \theta_j},
\ee
and we obtain\footnote{We emphasise this does not mean the mass
  matrix is $M^2 = a \otimes a$: simply that terms may receive an enhance
  by factors appearing in the exponential.}
\be
M^2_{ij} \sim a_i a_j V_{min},
\ee
with $V_{min}$ the magnitude of the potential at the AdS minimum. In KKLT, $D_{T_i} W = 0$ for all $i$
and $V_{min} \sim - 3 \frac{\vert W_0 \vert^2}{\mc{V}^2}$. As there are
$h^{1,1}$ independent phases, there is no reason for $M^2_{ij}$ to be degenerate and
we expect all eigenvalues to be $\mc{O} ( a^2 V_{min})$, where $a$ is the typical magnitude of the $a_i$.

The determination of physical masses also requires the K\"ahler potential. In general there is no explicit expression for the
overall volume $\mc{V}$ in terms of 4-cycle volumes $\tau_i$. The K\"ahler metric may however be written as \cite{hepth0403067}
\be
\label{GeneralKahlerP}
\mc{K}_{i \bar{j}} = \frac{G_{i \bar{j}}^{-1}}{\mc{V}^2}, \qquad \qquad
G_{i \bar{j}} = -\frac{3}{2} \left( \frac{k_{ijk}v^k}{\mc{V}} - \frac{3}{2} \frac{k_{imn}t^m t^n k_{jpq} t^p t^q}{\mc{V}^2} \right).
\ee
If $\mc{V} \sim \textrm{ (a few)} l_s^6$, $\mc{K}_{i \bar{j}} \sim \mc{O}(1)$ and the mass matrix $M^2_{ij}$
gives a good estimate of the scale of axion masses. If $\mc{V} \gg l_s^6$,
then $\mc{K}_{i \bar{j}} \ll \mc{O}(1)$, and $M^2_{ij}$ underestimates the axion masses.
$M^2_{ij}$ could only overestimate the axion masses if $\mc{V} \ll 1$.
This realm of moduli space is not accessible in a controlled fashion and
we do not concern ourselves with it.

In units where $M_P = 1$, we therefore have
\be
m_{\tau_i} \sim m_{c_i} \sim a_i \sqrt{V_{min}} \sim \frac{a_i W_0}{\mc{V}}.
\ee
The axion masses are consequently set by the value of the tree-level superpotential
$W_0$. This also determines the vacuum
energy and, implicitly, the energy scale of supersymmetry breaking
required to cancel the vacuum energy.
TeV-scale soft terms require hierarchically small $W_0$. For the
(gravity-mediated) case
studied in \cite{hepth0503216}, this required $W_0 \sim 10^{-13}$, with
\be
m_{\tau_i} \sim m_{a_i} \sim m_{3/2} \sim 10 \textrm{TeV}.
\ee
This scale is vastly greater than that associated with QCD instanton effects, and thus the axions
are incapable of solving the strong CP problem.
We could insist on a QCD axion, and require that $W_0$ be sufficiently small
that QCD instantons
dominate over the D-instanton effects of moduli stabilisation.
This would require $W_0 \sim 10^{-40}$. However, this scenario is entirely
excluded as the size moduli are light enough to violate fifth force experiments and the susy breaking scale would be $\mc{O}(10^{-14}
\textrm{eV})$.
Consequently, in the simplest KKLT scenario it is impossible to generate a QCD axion. The D3 axion
receives a high scale mass from fluxes, whereas the instanton effects
give the D7 axions
large masses comparable to the size moduli.

A similar argument holds for the exponentially large volume compactifications. As in KKLT, the D3 axion
receives a large flux-induced mass. The `small' cycles are stabilised by instanton effects, and these give the
corresponding axions masses of a similar scale to the size moduli, $m_{a_i} \sim m_{\tau_i} \sim m_{3/2}$.
One difference is that there is a modulus, the `large' modulus $\tau_b$, which
need not appear in the superpotential. It is stabilised through the
K\"ahler potential, and while it is massive its axionic partner indeed
remains massless.
However, this cycle is exponentially large, and
any gauge group supported on this cycle is far too weakly coupled to be QCD.
The same conclusion holds: the simplest version of this scenario does not generate a QCD axion.

The above formulates the `modulus anti-stabilisation problem': naive
scenarios of moduli stabilisation are incompatible with a QCD axion.

We next examine an apparent solution to this problem that in fact has
a subtle flaw.
We want a way to stabilise moduli without stabilising the axions.
Axions correspond to phases in the superpotential and do not appear in
the K\"ahler potential. If we included a multi-exponential term
$e^{-\alpha^i T_i}$ in the
superpotential, a massless axion would certainly survive,
as at least one phase would be absent.
As the size moduli all appear in the K\"ahler potential,
by solving the F-term equations we may hope to stabilise the size
moduli while leaving the axions massless.

To illustrate this idea, let us consider a toy model,
\bea
\label{ToyTorusK}
\mc{K} & = & - \ln (T_1 + \bar{T}_1) - \ln (T_2 + \bar{T}_2) - \ln
(T_3 + \bar{T}_3), \\
\label{LightAxionSuperpotential}
W & = & W_0 + Ae^{-2 \pi(T_1 + T_2 + T_3)}.
\eea
The K\"ahler potential (\ref{ToyTorusK}) is that appropriate for
toroidal orbifolds, with $\mc{V} = t_1 t_2 t_3$.
To understand where the superpotential could arise from, we can hypothesise that
the cycle (1+2+3) is the smallest cycle with only two fermionic
zero modes, and that instantons wrapping (for example) the cycle (2+3)
all have more than two zero modes and do not appear in the
superpotential. However, we are not here really concerned with the
microscopic origin of the superpotential: at this level we simply regard equations
(\ref{ToyTorusK}) and (\ref{LightAxionSuperpotential}) as defining the model.

The F-term equations $D_{T_1} W = D_{T_2} W = D_{T_3} W = 0$ give
\bea
\label{F1}
- 2 \pi A e^{-2 \pi (T_1 + T_2 + T_3)} - \frac{1}{T_1 + \bar{T}_1} (W_0 + A
e^{-2 \pi (T_1 + T_2 + T_3)}) & = & 0, \\
\label{F2}
- 2 \pi A e^{-2 \pi (T_1 + T_2 + T_3)} - \frac{1}{T_2 + \bar{T}_2} (W_0 + A
e^{-2 \pi (T_1 + T_2 + T_3)}) & = & 0, \\
\label{F3}
- 2 \pi A e^{-2 \pi (T_1 + T_2 + T_3)} - \frac{1}{T_3 + \bar{T}_3}
(W_0 + A e^{-2 \pi (T_1 + T_2 + T_3)}) & = & 0.
\eea
These immediately imply $\tau_1 = \tau_2 = \tau_3$ and equations
(\ref{F1}) to (\ref{F3}) collapse to
\bea
2 \pi A e^{-6 \pi \tau_1} e^{-2 \pi i(\theta_1 + \theta_2 + \theta_3)} +
\frac{1}{2 \tau_1} (W_0 + A e^{-6 \pi \tau_1} e^{-2 \pi i(\theta_1 +
  \theta_2 + \theta_3)}) & = & 0.
\eea
While the sum $\theta_1 + \theta_2 + \theta_3$ is fixed, there are
clearly two axionic directions not relevant for the solution of the
F-term equations. On the other hand, there is a unique value for the
size moduli such that the F-term equations are solved.
Except for the massless axionic directions, the scales of the masses
are unaltered from above, and we would expect
\be
m_{\tau_i} \sim m_{\theta_1 + \theta_2 + \theta_3} \sim
\frac{W_0}{\mc{V}}, \qquad
m_{\theta_1 - \theta_2} = m_{\theta_1 - \theta_3} = 0.
\ee
As this is supergravity rather than rigid supersymmetry, there is
no contradiction in having a mass splitting for the multiplet in the
presence of unbroken supersymmetry.

While this seems promising, there is in fact a
serious problem with the above. Even though all F-term
equations can be solved, numerical investigation shows that at the
supersymmetric locus the resulting scalar potential is
tachyonic, with signature $(+,-,-)$. Although supersymmetry ensures the
moduli are Breitenlohner-Freedman stable \cite{BreitenlohnerFreedman},
this notion of AdS stability ceases to be relevant
after the (necessary) uplift.

We now show that these tachyons are in fact generic for any attempt to stabilise the
moduli supersymmetrically while preserving unfixed axions.

\subsection{A No-Go Theorem}

We suppose we have an arbitrary $\mc{N} = 1$ supergravity with
moduli fields, $\Phi_{\alpha}$, $T_{\beta} = \tau_{\beta} + i b_{\beta}$, where the $b_{\beta}$ are the
axions. We write the superpotential and K\"ahler potential as
\bea
W & = & W(\Phi_{\alpha}, T_{\beta}), \\
\label{KKaxionK}
\mc{K} & = & \mc{K}(\Phi_{\alpha}, T_{\beta} + \bar{T}_{\beta}).
\eea
The Peccei-Quinn symmetry $b_{\beta} \to b_{\beta} + \epsilon_{\beta}$ implies the form
of $(\ref{KKaxionK})$ should hold in perturbation theory.

We further suppose we have solved
\be
\label{FFEqs}
D_{\Phi_{\alpha}} W = 0 \textrm{ and } \qquad D_{T_{\beta}} W = 0
\ee
for all $\alpha$ and $\beta$, but that at least one axion $b_u = \sum_{\beta}
\lambda_{\beta} b_{\beta}$ is unfixed: the solution to (\ref{FFEqs}) is
independent of $\langle b_u \rangle$.

We redefine the basis of chiral superfields so that there exists a
superfield $T_u$ with $b_u =
\textrm{Im}(T_u)$,
\bea
T_1 & \to & T_u, \nonumber \\
T_2 & \to & T_2, \nonumber \\
T_n & \to & T_n.
\eea
This is a good redefinition as it does not affect holomorphy
properties.

As the solution to all F-term equations is independent of $b_u$,
$b_u$ is a flat direction of the potential (\ref{FTermPot}) at the supersymmetric
locus.\footnote{The requirement of \emph{flatness} is stronger than
  the requirement that the axion simply be \emph{massless}. Flatness
  is the right requirement, as if an axion is fixed in any way it does
not solve strong CP.}
The potential at the supersymmetric locus is given by
\be
V = - 3 e^{\mc{K}} \vert W \vert^2.
\ee
As $b_u$ does not appear in $\mc{K}$, it follows that if $b_u$ is a
flat direction $\vert W \vert$
must be independent of $b_u$. Up to one exception this then implies that
$W$ is independent of $b_u$.

The sole exception is if
$b_u$ purely represents an overall phase, i.e. $W = e^{-a T_u}$ with
no constant term. In a IIB context, this may potentially arise if
the flux superpotential exactly vanishes due to a discrete symmetry \cite{hepth0411061},
while a combination of non-perturbative effects and K\"ahler
corrections stabilise the K\"ahler moduli. While potentially
interesting, this is an exceptional case and we do not analyse it further.

If $W$ has no explicit dependence on $b_u = \textrm{Im}(T_u)$, it
follows by holomorphy that it also has no explicit dependence on
$\tau_u = \textrm{Re}(T_u)$ and hence on $T_u$. Therefore
\be
\partial_{T_u} W \equiv 0.
\ee
However, as $D_{T_u} W = 0$, it follows that at the supersymmetric locus,
$$
\textrm{either } \qquad (\partial_{T_u} K) = 0 \qquad \textrm{ or }
\qquad W = 0.
$$
The latter is overdetermined and non-generic, so we first focus on
$(\partial_{T_u} K) = 0$.

Direct calculation now shows that the $\tau_u$ direction is tachyonic
at the supersymmetric locus. To see this, note that from the scalar potential
(\ref{FTermPot})
\be
\label{VDeriv}
\partial_{\tau_u} V = e^{\mc{K}} \mc{K}^{i \bar{j}} \left( \partial_{\tau_u} (D_i W) D_{\bar{j}}
\bar{W} + D_i W \partial_{\tau_u} (D_{\bar{j}} \bar{W}) \right) - 3
(\partial_{\tau_u} K) e^K W \bar{W}.
\ee
We have used $\partial_{\tau_u} W \equiv 0$
and have only kept terms that
will give non-vanishing contributions to $\partial_{\tau_u}
\partial_{\tau_u} V$ at the supersymmetric locus.
Expanding $D_i W$ and again using $\partial_{\tau_u} W \equiv 0$,  (\ref{VDeriv})
simplifies to
\be
\partial_{\tau_u} V = e^{\mc{K}} \mc{K}^{i \bar{j}} \left( \partial_{\tau_u} (\partial_i \mc{K}) W
(D_{\bar{j}} \bar{W}) + D_i W \partial_{\tau_u}(\partial_{\bar{j}} \mc{K}) \bar{W} \right) - 3
(\partial_{\tau_u} \mc{K}) e^{\mc{K}} W \bar{W}.
\ee
If we again only keep terms non-vanishing at the supersymmetric locus,
the second derivative is
\be
\partial_{\tau_u} \partial_{\tau_u} V = e^{\mc{K}} \mc{K}^{i \bar{j}} \left(
2 \partial_{\tau_u} (\partial_i \mc{K}) \partial_{\tau_u} (\partial_{\bar{j}} \mc{K}) W
\bar{W} \right) - 3 (\partial_{\tau_u} \partial_{\tau_u} \mc{K}) e^{\mc{K}} W \bar{W}.
\ee
Now, as $\tau_u = \half (T_u + \bar{T}_u)$,
$$
\partial_{\tau_u} \mc{K}(T + \bar{T}) =
2 \partial_{T_u} \mc{K}(T + \bar{T}),
$$
and we have
\bea
\partial_{\tau_u} \partial_{\tau_u} V  & = &
4 e^{\mc{K}} W \bar{W} ( 2 \mc{K}^{i \bar{j}} \mc{K}_{i u} \mc{K}_{u \bar{j}} - 3 \mc{K}_{u \bar{u}} )
\nonumber \\
\label{TachyonMass}
& = & - 4 e^{\mc{K}} W \bar{W} \mc{K}_{u \bar{u}},
\eea
where we have used $\mc{K}_{i \bar{j}} = \mc{K}_{ij}$.
As $\mc{K}_{i\bar{j}}$ is a metric, $\mc{K}_{u \bar{u}}$ is positive definite and it follows
that the $\tau_u$ direction is tachyonic.

Now consider the $W = 0$ case. As indicated above, this is
non-generic: even if $W$ originally vanishes, it is expected
to receive non-perturbative corrections which make it non-vanishing.
Even so, it follows easily that
\be
\partial_{\tau_u} \partial_{\tau_u} V = 0,
\ee
and so the $\tau_u$ size modulus is massless, leading to unobserved fifth forces.

The above gives a no-go theorem: there does not exist any supersymmetric minimum of the F-term
potential consistent with stabilised moduli and unfixed axions.

It is in the nature of no-go theorems that they admit loopholes, so
let us discuss ways around this result. One point to consider is the form of the
K\"ahler potential, as we have used in the above argument the fact that
\be
\mc{K} = \mc{K}(T + \bar{T}).
\ee
While true in perturbation theory because of the axionic shift symmetry, this equation will break down
nonperturbatively, and the argument showing that the $\tau_u$
direction is tachyonic will no longer hold. However, this same breakdown will cause
$\mc{K}$, and hence the potential $V$, to depend on the axion.
As this lifts the required axionic flat direction, the no-go theorem will cease to apply.

A second loophole is that although the F-term potential might be
tachyonic, the D-term potential might come to the rescue. For example,
a Fayet-Iliopoulos term might have exactly the right structure to render the
supersymmetric locus an actual minimum of the full potential. However
this seems implausible in the presence of many tachyonic
directions. A similar approach would be to try and set $W=0$ and then
rely entirely on D-terms to stabilise the moduli.
Another possibility (discussed recently in
\cite{hepth0602120}) is that an anomalous U(1) might remove the
tachyonic directions. While the massive gauge boson will eat the
axionic degree of freedom, an axionic direction may survive in the
phase of a scalar charged under the U(1).

A third loophole is that stability might not rely on the existence of
an actual minimum for the potential.
Equation (\ref{TachyonMass}) involves the K\"ahler metric $\mc{K}_{u
  \bar{u}}$. The kinetic term for $\tau_u$ is $\mc{K}_{u \bar{u}}
\partial_\mu \tau_u \partial^\mu \tau_u$. If we just consider the
$\tau_u$ direction, the physical mass is therefore
\be
m_{\tau_u}^2 = -2 e^{\mc{K}} W \bar{W} = -\frac{8}{9} \vert m_{BF} \vert^2,
\ee
where $m_{BF}$ is the relevant
Breitenlohner-Freedman bound. As tachyonic modes can be stable in AdS,
one could argue that it is sufficient simply to solve the F-term
equations and not to worry about whether the resulting locus is an
actual minimum of the potential.

While this point is more substantial, it does not resolve the
problem. The real world is not AdS, and for stability requires a positive definite
mass matrix. For any realistic model, the vacuum energy must be
uplifted such that it vanishes.
After this uplift,
the extra geometric advantages of AdS go away and the tachyons can no
longer be supported. As there may be many tachyons present - one for
each massless axion -
the entire problem of moduli stabilisation must necessarily be solved over again in the uplifting.
As the uplift is generally the least controlled part of the procedure,
this seems a hard problem.
While the uplift \emph{may} remove the tachyons, at the present level
of understanding this seems pure hypothesis.
It is
then very unclear how useful the original supersymmetric AdS saddle point is, and whether
it is a suitable locus to uplift.

The argument above suggests that supersymmetric solutions are unpromising
starting points from which to address the strong CP problem. Either
all moduli appear in the superpotential, in which case there is no
light axion, or a modulus is absent from the superpotential, in which
case the potential is tachyonic.
The fourth and most obvious loophole is then to give up on the requirement
of supersymmetric minima, and search for nonsupersymmetric minima of
the potential with massless axions.

We shall consider this point in the next section. Before we do so we
also observe that while
our focus here is on axions, the argument above is also
relevant for moduli stabilisation. For example, in the weakly
coupled heterotic string, the one-loop corrections to the gauge
kinetic function (\ref{HetOneLoop}) imply that gaugino
condensation generates a superpotential
\be
\label{HetSuperpotential}
W_{n.p.} = A e^{-\alpha S + \beta_i T^i}.
\ee
It has been proposed to use the superpotential
(\ref{HetSuperpotential}), together with a constant term $W_0$, to
stabilise the dilaton and K\"ahler moduli by solving $D_S W = D_{T^i} W =
0$.
However, there is only one phase - and hence only one axion -
explicitly present in (\ref{HetSuperpotential}). The above
argument shows that the resulting scalar potential will actually be
tachyonic at the supersymmetric locus, with signature $(+,-,\ldots,-)$.

A similar result will apply to the recent study of IIA flux compactifications
with all NSNS and RR fluxes turned on. In this context it is also
found that the solution of the F-term equations is independent of many
of the axions present (those associated with the $C_3$ field).
The above implies that as long as $h^{2,1} \neq 0$ the
supersymmetric locus is always tachyonic, with one tachyon present for
every massless axion. Tachyons have been recognised in particular
models \cite{hepth0602120, hepth0505160}, but from the above they
would seem to be very generic.

\subsection{Non-supersymmetric Minima with Massless Axions}
\label{secnonsusyminima}

The above no-go theorem shows that supersymmetric moduli stabilisation
is not a good starting point from which to solve the strong CP problem.
This implies we ought to consider non-supersymmetric moduli stabilisation.
That there is no no-go theorem for non-supersymmetric minima with
massless axions can be shown by construction: for example, the exponentially large
volume compactifications of \cite{hepth0502058, hepth0505076} all
contain a massless axion associated with the large cycle controlling
the overall volume. As indicated above, this cannot be a QCD axion, as
any brane on this cycle is very weakly coupled. If we want to try and
force this cycle into being a QCD axion, we can tune the
parameters to force the minimum of this potential to relatively
small volumes. That this is possible can be confirmed numerically.
Another possibility would be a purely perturbative stabilisation of
the volume modulus, solely using K\"ahler corrections (in which axions
do not appear). This has been discussed in \cite{hepth0507131, hepth0508171}, although
without an explicit example.

We shall not dwell on these possiblities. First, because the resulting
axion decay constant would be, as we shall see in section
\ref{AxionDecayConstantSec}, close to the Planck scale and outside the
allowed window and secondly, because at such small volumes there is no
good control parameter. As we require non-supersymmetric minima we
shall
base our discussion around the large-volume
models of \cite{hepth0502058, hepth0505076}. For now we only discuss
moduli stabilisation but in
section \ref{AxionDecayConstantSec} we shall show that
these can also realise phenomenological values for $f_a$.

We illustrate the discussion with a three-modulus toy model, in which we assume the volume may be
expressed in terms of 4-cycles as
\be
\label{VolExpression}
\mc{V} = (T_1 + \bar{T}_1)^{\frac{3}{2}} - (T_2 +
\bar{T}_2)^{\frac{3}{2}} - (T_3 + \bar{T}_3)^{\frac{3}{2}}.
\ee
We use three moduli as this turns out to be the minimal number
required for our purposes: clearly, this is no significant restriction.
Expressed in terms of 2-cycles, (\ref{VolExpression}) corresponds to
\be
\label{Toy2TI}
\mc{V} = \lambda( t_1^3 - t_2^3 - t_3^3).
\ee
We note (\ref{Toy2TI})
satisfies the requirement that $\frac{\partial^2 \mc{V}}{\partial t_i \partial t_j}$ have signature $(+,-,-)$.
We may perhaps think of this toy model as a $\mbb{P}^3$ with two points blown up.
Denoting the cycles by 1, 2 and 3, the K\"ahler potential is
\be
\label{ToyK}
\mc{K} = - 2 \ln \left(  (T_1 + \bar{T}_1)^{\frac{3}{2}} - (T_2 +
\bar{T}_2)^{\frac{3}{2}} - (T_3 + \bar{T}_3)^{\frac{3}{2}} \right) -
\frac{\xi}{g_s^{3/2} \mc{V}},
\ee
where we have included the leading large-volume behaviour of the
$\alpha'^3$ correction of \cite{hepth0204254}. $g_s$ is fixed by the
fluxes and in (\ref{ToyK}) should be regarded as a tunable parameter.
For superpotential, we shall take
\be
W = W_0 + e^{-\frac{2 \pi}{n} \left( T_2 + T_3 \right)}.
\ee
This could arise from gaugino condensation on a stack of $n$ branes
wrapping the combined cycle 2+3.\footnote{The use of gaugino
  condensation rather
  than instanton effects is necessary to ensure that the cycle 2+3 is
  large enough to contain QCD.} QCD will be realised as a stack of branes
wrapping cycle 3.

In the limit $\mc{V} \gg 1$ with $\tau_2$ and $\tau_3$ small,
the leading functional form of the scalar potential is (omitting numerical factors)
\be
\label{ToyScalarPotential}
V = \frac{(\sqrt{\tau_2} + \sqrt{\tau_3})e^{-\frac{2 \pi}{n} 2(\tau_2 +
    \tau_3)}}{\mc{V}}
- \frac{(\tau_2 + \tau_3) e^{-\frac{2 \pi}{n} (\tau_2 +
    \tau_3)}}{\mc{V}^2}
+ \frac{\xi}{g_s^{3/2} \mc{V}^3}.
\ee
The minus sign in (\ref{ToyScalarPotential}) arises from minimising the potential
for the axion $\textrm{Im}(T_2 + T_3)$. The axions $\textrm{Im}(T_1)$ and $\textrm{Im}(T_2 -
    T_3)$ do not appear in (\ref{ToyScalarPotential}) and are unfixed.
By considering the limit $\mc{V} \to \infty$, $\frac{2 \pi (\tau_2 + \tau_3)}{n} \sim
    \ln \mc{V}$, it follows that as $\mc{V} \to \infty$ the potential
    (\ref{ToyScalarPotential}) goes to zero from below. As by
    adjusting $g_s$ we can make the third term of
    (\ref{ToyScalarPotential}) arbitrarily large, we can ensure the
    potential remains positive until arbitrarily large volumes, and
    thus any minimum will be at exponentially large volumes.

Is there a minimum? The potential is clearly symmetric under $\tau_2 \leftrightarrow \tau_3$, and the potential
restricted to the locus $\tau_2 = \tau_3$ indeed has a minimum at
exponentially large volumes. Because of the symmetry $\tau_2
\leftrightarrow \tau_3$, this `minimum' is also a
critical point of the full potential. However, it is not a minimum of
the full potential. At fixed $\tau_2 + \tau_3$ and fixed $\mc{V}$,
(\ref{ToyScalarPotential}) depends only on $\sqrt{\tau_2} +
\sqrt{\tau_3}$. For fixed $\tau_2 + \tau_3$, this is \emph{maximised}
at $\tau_2 = \tau_3$, and the mode
$\tau_2 - \tau_3$ is tachyonic at this locus. We have not
investigated whether this tachyon satisfies the Breitenlohner-Freedman
bound for AdS stability.
This is for
the same reasons as above: once we uplift, the geometric protection of
AdS ceases to be relevant. Consequently the fields in
(\ref{ToyScalarPotential}) run away either to
$\tau_2 = 0$ or $\tau_3 = 0$,
where one of the blow-up cycles collapses.

This result shows that the
above toy model does not, by itself, have a minimum of the potential with
a massless QCD axion. We may ask whether this is a feature of the
geometric details of the model - for example,
whether a different choice of triple intersection form in
(\ref{Toy2TI}) would alter this result. We have investigated several other
toy models without finding a minimum, and while we have no proof we
suspect none exists so long as the K\"ahler potential is given by (\ref{ToyK}).

This is bad news, but it is controllable bad news.
The instability above is very
particular: there is no instability either for the overall volume or for
the sum of the blow-up volumes $\tau_1 + \tau_2$,
but only for the difference $\tau_1 - \tau_2$. The effect of the
instability is to drive one
of the blow-up cycles
to collapse. Consequently, the instability can be cured by \emph{any}
effect that becomes important at small cycle
volume and prevents
collapse.

For example, the addition of a term
\be
\label{NewTerm}
\frac{1}{\sqrt{\tau_2} \mc{V}^3} + \frac{1}{\sqrt{\tau_3} \mc{V}^3}
\ee
to (\ref{ToyScalarPotential}) would obviously stabilise the cycles
$\tau_2$ and $\tau_3$ against collapse and generate a minimum of the
potential.
As this term does not affect the
argument that in the $\mc{V} \to \infty$ limit the potential approaches
zero from below, the resulting minimum would still be at exponentially
large volume.

We discuss in greater detail in the appendix which K\"ahler
corrections are and are not allowed. For now we note
that terms of the form (\ref{NewTerm}) may be generated from a correction to the K\"ahler potential,
\be
\label{KahlerCorrectionNewTerm}
\mc{K} + \delta \mc{K} = - 2 \ln (\mc{V}) + \frac{\epsilon
  \sqrt{\tau_2}}{\mc{V}} + \frac{\epsilon \sqrt{\tau_3}}{\mc{V}}.
\ee
For simplicity we have kept the $\tau_2 \leftrightarrow \tau_3$
symmetry.
Such a correction is motivated by the fact that it gives corrections
to the
K\"ahler metrics $\mc{K}_{2 \bar{2}}$ and
$\mc{K}_{3 \bar{3}}$ suppressed by
factors of $g^2$ for the field theory on the relevant cycle. More specifically,
\be
\mc{K}_{2 \bar{2}} + \delta \mc{K}_{2 \bar{2}} =
\frac{3}{2\sqrt{2 \tau_2} \mc{V}}\left( 1 - \frac{\epsilon}{12 \sqrt{2} \tau_2} \right).
\ee
As $\tau_2 = \frac{1}{g^2}$ for a brane wrapping the cycle 2, the
  correction is
suppressed by $g^2$.

The inverse metric involves an infinite series of terms diverging in
the $\tau_2 \to 0$ and $\tau_3 \to 0$ limit. For example,
\be
\label{InverseKDivergence}
\mc{K}^{2 \bar{2}} = \frac{2 \sqrt{2 \tau_2} \mc{V}}{3} \left( 1 +
\frac{\epsilon}{12 \sqrt{2} \tau_2} + \frac{\epsilon^2}{288 \tau_2^2} +
\ldots \right).
\ee
This is easy to understand: at $\tau_2 = \frac{\epsilon}{12
  \sqrt{2}}$, the K\"ahler metric $\mc{K}_{2 \bar{2}}$ goes to zero
and the inverse metric diverges. This divergence can be seen by resumming
(\ref{InverseKDivergence}). In the physical potential, this divergence
will create a positive wall at finite values of $\tau_2$ and
$\tau_3$. The positivity can be seen from the fact that the divergence
in $\mc{K}^{-1}$ will only appear in the term
\be
e^{\mc{K}} \mc{K}^{i \bar{j}} D_i W D_{\bar{j}} \bar{W},
\ee
which is manifestly positive
definite. Consequently the potential will diverge positively
at finite values of $\tau_2$ and $\tau_3$, and so a stable minimum
must exist for both $\tau_2$ and $\tau_3$.

We have now outlined, in the context of the scenario of
\cite{hepth0502058, hepth0505076}, a way to stabilise all the size
moduli while keeping a massless QCD axion left over. We would like
cycle 3 to support QCD:
by adjusting $\epsilon$, we can always make the correction
(\ref{KahlerCorrectionNewTerm})
sufficiently large to ensure that $\tau_3$ is stabilised with
the correct size for QCD. For intermediate string scales, this
requires $\tau_3 \sim 10$. As the actual correction would be very hard
to calculate, at this level we just adjust $\epsilon$
phenomenologically.
Of course, the complexity of a real model is much greater than that of
(\ref{KahlerCorrectionNewTerm}).
However we note again that, even though the corrections cannot be calculated, our
proposal for moduli stabilisation only requires that they exist and come with the right sign to prevent collapse.

While the above has been with a toy example, the above
approach will apply to any model in which the moduli are stabilised
along the lines of \cite{hepth0502058, hepth0505076}. Keeping an axion
massless introduces an instability causing a blow-up cycle to want to
collapse. K\"ahler corrections that become important at small volume
can stabilise this cycle but will not affect the overall structure of
the potential, and in particular will not affect the stabilisation of
the volume at $\mc{V} \gg 1$.

\subsection{Higher Instanton Effects in the Axion Potential}
\label{sechigherinstanton}

We have given above a K\"ahler potential and superpotential that will
stabilise the moduli while containing a candidate QCD axion.
The nonperturbative terms in the superpotential are in general
just the leading terms in an instanton expansion. Even though the higher order terms
may be highly suppressed and irrelevant to moduli stabilisation, they could still
lift the flat direction associated with the massless axion.
Given that $\Lambda_{QCD} \ll M_P$, even highly subleading terms could dominate over QCD instantons.

Let us estimate the general magnitude of such instanton effects.
The magnitude of brane instantons depends on the volumes of the cycles they can wrap. Generally there will be many such cycles,
whose sizes depend on the stabilised moduli, but
minimally there must always exist the cycle which support the QCD stack. It follows from the DBI action
that the gauge coupling for a D7-brane stack is\footnote{v3: This corrects a factor of 2 from earlier versions. To deduce the gauge
coupling from reduction of the DBI action, it is necessary to ensure the gauge group generators are normalised according to the
phenomenology conventions $Tr(T^\alpha T^\beta) = \half \delta^{\alpha \beta}$.}
$$
\frac{1}{g^2} = \frac{\textrm{Re}(T)}{4 \pi} \Rightarrow \alpha^{-1} =
\frac{4 \pi}{g^2} = \textrm{Re}(T) = \tau.
$$
This defines the gauge coupling at the high scale where the effective field theory becomes valid.
When $m_s \sim M_P$, this is in essence the string scale, but if $m_s \ll M_P$,
the difference between $m_{KK}$ and $m_s$ become significant. It is a
subtle issue whether $m_s$ or $m_{KK}$ is the appropriate high
scale. If QCD is supported on a small cycle within a large internal
space, the KK modes associated with the bulk will be uncharged under
QCD and will not contribute to the running coupling. KK modes of the
QCD cycle will contribute, but these will be at masses comparable to
the string scale. In considering
the running coupling, we therefore use $m_s$ as the high scale rather
than $m_{KK}$.
We consider a wide range of string scales and take a sampling of high scale values from
$10^8 \to 10^{16}$ GeV. Given a string scale, the internal volume is determined by $m_s
\sim \frac{M_P}{\sqrt{\mc{V}}}$.

The QCD coupling runs logarithmically with energy scale, with
$$
\alpha_{QCD}^{-1}(10^2 \textrm{GeV}) \sim 9 \qquad \textrm{ and } \qquad
\alpha_{QCD}^{-1}(10^{16} \textrm{GeV}) \sim 25.
$$
The required high-scale couplings and cycle sizes are given in table \ref{Table1},
together with the action for a D3-brane instanton wrapping the same cycle as the QCD stack.
Its magnitude is set by $\sim e^{-2 \pi T}$ and
we show in table \ref{Table1} the approximate magnitude of single- and double-instanton effects.
In addition to the QCD cycle, there may be other cycles which instantons
may wrap. We do not include these for two reasons: first,
whether such instantons would generate a potential for the QCD axion
is model-dependent\footnote{It seems odd that such instantons could
  affect the QCD axion at all. However, if QCD is on cycle 3, and the
  axion $b_2 + b_3$ is fixed by effects on cycle 2+3, an instanton
  solely on cycle 2 effectively generates a potential for the QCD axion.}, and secondly, we can always arrange the
model such that the QCD cycle is smallest and hence
dominates the instanton expansion.
We observe that the magnitude of the required cycle volume, and thus the magnitude of potential instanton effects,
varies significantly with the string scale.
\begin{table}
\caption{Cycle sizes and instanton amplitudes for various UV scales}
\label{Table1}
\centering
\vspace{3mm}
\begin{tabular}{|c|c|c|c|c|c|}
\hline
$\textrm{E}_{UV}$ & $10^8$ GeV & $10^{10}$ GeV & $10^{12}$ GeV & $10^{14}$ GeV & $10^{16}$ GeV\\
\hline
$\alpha_{QCD}^{-1}(\textrm{E}_{UV})$ & 15.8 & 18.1 & 20.4 & 22.7 & 25 \\
\hline
\textrm{Re}(T) = $\alpha^{-1}$ & 15.8 & 18.1 & 20.4 & 22.7 & 25 \\
\hline
$e^{-2 \pi T}$ & $7.7 \ti 10^{-44}$ & $2 \ti 10^{-50}$ & $2 \ti 10^{-56}$ & $6 \ti 10^{-63}$ & $6 \ti 10^{-69}$ \\
\hline
$e^{-4 \pi T}$ & $5.9 \ti 10^{-87}$ & $4 \ti 10^{-100}$ & $4 \ti 10^{-112}$ & $3.6 \ti 10^{-127}$ & $3.6 \ti 10^{-139}$ \\
\hline
\end{tabular}
\end{table}
If present, such D-instantons would generate a potential for the QCD axion. To compare their magnitude to that of QCD effects,
we need an estimate of their contribution to the scalar potential. In this context we only care
about terms containing a phase and so contributing to the axion potential.
To this end, the relevant term from the scalar potential (\ref{NoScaleScalarPotential}) is
\be
V_{axion} = e^{\mc{K}} \left( \mc{K}^{i \bar{j}} \left( \partial_i W (\partial_{\bar{j}} \mc{K}) \bar{W} + c.c. \right) \right).
\ee
A superpotential instanton contribution $e^{-2 \pi n  T_i}$ generates a term
\be
\label{AxionPotential}
V_{axion} = \frac{- 2 a_i \tau_i W_0}{\mc{V}^2} e^{-2 \pi n \tau_i} \cos(\theta_i).
\ee
The absolute magnitude of (\ref{AxionPotential}) depends on the value
of $n$, the internal volume $\mc{V}$
and the tree-level
superpotential $W_0$. As we are looking towards phenomenology
we also take $m_{3/2} = \frac{W_0}{\mc{V}} \sim 1 \textrm{TeV} \sim 10^{-15} M_P$,
as appropriate for gravity-mediated TeV-scale soft terms.
Note that for models built around the KKLT scenario, we always have $m_s \gtrsim M_{GUT}$ and
only the largest value of $\textrm{E}_{UV}$ is achievable.
In table \ref{Table2} we give the internal volumes required for
each UV scale, as well as the resulting absolute magnitude of 1-,2- and 3-instanton superpotential corrections to the scalar potential.
For the same reasons as above, we only consider instantons wrapping the QCD cycle.
\begin{table}
\caption{Magnitude of Axion Potentials from Superpotential Instanton Effects}
\label{Table2}
\centering
\vspace{3mm}
\begin{tabular}{|c|c|c|c|c|c|}
\hline
$\textrm{E}_{UV}$ & $10^8$ GeV & $10^{10}$ GeV & $10^{12}$ GeV & $10^{14}$ GeV & $10^{16}$ GeV\\
\hline
$\mc{V}$ & $10^{20}$ & $10^{16}$ & $10^{12}$ & $10^{8}$ & $10^4$ \\
\hline
$V_{\textrm{1-instanton}}$ & $10^{-79} M_P^4$ & $10^{-81} M_P^4$ & $10^{-83} M_P^4$ & $10^{-85} M_P^4$ & $10^{-87} M_P^4$ \\
\hline
$V_{\textrm{2-instanton}}$ & $10^{-122} M_P^4$ & $10^{-131} M_P^4$ & $10^{-139} M_P^4$ & $10^{-151} M_P^4$ & $10^{-155} M_P^4$ \\
\hline
\end{tabular}
\end{table}

As well as superpotential effects, there are also nonperturbative corrections to the K\"ahler potential (the
perturbative corrections to $\mc{K}$ do not have an axionic dependence).
While smaller, these are easier
to generate - the instantons can have four fermionic zero modes rather than only two. A correction
\be
\mc{K} = - 2 \ln (\mc{V}) \to \mc{K} = -2 \ln (\mc{V} + e^{-2 \pi n T})
\ee
will generate effects in the scalar potential at order
$$
V_{\delta \mc{K}} \sim \frac{W_0^2}{\mc{V}^3} e^{-2 \pi n T}.
$$
and thus generate a potential for the QCD axion $\theta$ of the form
$$
V_{\delta K} \cos (\theta + \alpha).
$$
Again assuming TeV-scale (visible)
SUSY breaking, $\frac{W_0}{\mc{V}} \sim 10^{-15}$, the magnitudes of such effects are shown in table \ref{Table3}.
\begin{table}
\caption{Magnitude of Axion Potentials from K\"ahler Potential Instanton Effects}
\label{Table3}
\centering
\vspace{3mm}
\begin{tabular}{|c|c|c|c|c|c|}
\hline
$\textrm{E}_{UV}$ & $10^8$ GeV & $10^{10}$ GeV & $10^{12}$ GeV & $10^{14}$ GeV & $10^{16}$ GeV\\
\hline
$V$ & $10^{20}$ & $10^{16}$ & $10^{12}$ & $10^8$ & $10^4$ \\
\hline
$V_{\textrm{1-instanton}}$ & $10^{-99} M_P^4$ & $10^{-97} M_P^4$ & $10^{-95} M_P^4$ & $10^{-93} M_P^4$ & $10^{-91} M_P^4$ \\
\hline
$V_{\textrm{2-instanton}}$ & $10^{-143} M_P^4$ & $10^{-147} M_P^4$ & $10^{-151} M_P^4$ & $10^{-155} M_P^4$ & $10^{-159} M_P^4$ \\
\hline
\end{tabular}
\end{table}

 The axion potential originating from QCD effects and relevant to the
strong CP problem is
$$
V_{QCD} \sim \Lambda_{QCD}^4 (1 - \cos(\theta)),
$$
with $\Lambda_{QCD} \sim 2 \ti 10^{-19} M_P$ and $\Lambda_{QCD}^4 \sim 10^{-75} M_P^4$. We require QCD effects to be sufficiently
dominant to be consistent with the failure to observe CP violation in strong interactions.
Suppose we have a potential
\be
V = A (1 - \cos(\theta)) + \epsilon \cos(\theta + \gamma).
\ee
If $A \gg \epsilon$, the minimum is displaced from $\theta = 0$ by
$\delta \theta \sim \frac{ \epsilon }{A}$. Observationally, $\vert \theta \vert < 10^{-10}$,
and thus non-QCD contributions must have absolute magnitude smaller than $10^{-85} M_P^4$.
By comparison with tables $\ref{Table2}$ and $\ref{Table3}$
it follows that in order for QCD instantons to dominate the axion
potential, the instanton corrections to the 
K\"ahler potential are not constrained,
while the 1-instanton superpotential correction may or may not be
present depending on factors of $2 \pi$ and the precise value of the string
scale. Higher instanton corrections to the superpotential are not constrained.

To contribute to a superpotential (K\"ahler potential) an instanton must have at most 2 (4) fermionic zero modes, to generate
$\int d^4x d^2 \theta$ and $\int d^4 x d^2 \theta d^2 \bar{\theta}$ terms respectively. In the absence of flux, there is a
necessary condition on a divisor to generate a superpotential
\cite{hepth9604030}: it must have holomorphic
Euler characteristic one, $\chi_g(D) = 1$.
In the presence of flux, this condition may be relaxed. The number of zero modes on an instanton, and thus
its ability to appear in either the K\"ahler or superpotential,
may also be affected by the presence of the stack of QCD branes wrapping the would-be instanton cycle.
We shall not attempt to analyse this question for specific `real' models, but
by fiat simply assume the necessary instantons to be absent from
the potential. In this regard it is encouraging that the number of
instantons required to be suppressed is quite limited.

\section{The Axion Decay Constant}
\label{AxionDecayConstantSec}

\subsection{Magnitude}

The last section was devoted to solving the strong CP problem: keeping
an axion light while stabilising the moduli.
However, even achieving this does not resolve all phenomenological problems.
Given that a QCD axion exists, as indicated
in the introduction there are strong bounds
on the axion decay constant $f_a$ of equation (\ref{axionLagrangian}):
$10^9 \textrm{GeV} < f_a < 10^{12} \textrm{GeV}$.
The lower bound, from supernova cooling, is hard. While the upper bound may
be relaxed by considering non-standard cosmologies, here we shall also
treat this as hard.
We want to estimate $f_a$ in some moduli stabilisation scenarios.
In IIB compactifications, the axionic coupling to QCD arises from the clean and model-independent
Chern-Simons coupling.
However, to obtain the physical value of $f_a$ the axion must be
canonically normalised.
This depends on the K\"ahler metric,
and in particular on where the moduli are stabilised.

We assume we can write the K\"ahler potential as
\be
\mc{K} = \mc{K}(T_i + \bar{T}_i),
\ee
with $\mc{K}$ real.
This is true in perturbation theory, owing to the axionic shift
symmetry,
and any nonperturbative violations are small enough to be
irrelevant for this purpose.
In this case the kinetic terms for the axionic and size moduli do not mix.
Noting that $\mc{K}_{i \bar{j}} = \mc{K}_{j \bar{i}}$, we have for any $i$ and $j$
\bea
& & \mc{K}_{i \bar{j}} ( \partial_\mu T^i \partial^\mu \bar{T}^j ) +
\mc{K}_{j \bar{i}} ( \partial_\mu T^j \partial^\mu \bar{T}^i ) \nonumber \\
& = & \mc{K}_{i \bar{j}} \left( (\partial_{\mu} \tau_i + i \partial_\mu c_i)(\partial^\mu \tau_j -i \partial^\mu c_j)
+ (\partial_{\mu} \tau_j + i \partial_\mu c_j)(\partial^\mu \tau_i -i \partial^\mu c_i) \right) \nonumber \\
& = & \mc{K}_{i \bar{j}} (2 \partial_\mu \tau_i \partial^\mu \tau_j + 2 \partial_\mu c_i \partial^\mu c_j),
\eea
and the two sets of terms decouple.

Let us first show that if both the overall volume and the cycle
volumes are comparable to the string scale, then as expected $f_a \gtrsim 10^{16}
\textrm{GeV}$. Suppose an axion $c_i$ is to be the QCD axion. The
Lagrangian for this axion is
\be
\mc{K}_{i \bar{i}} \partial_\mu c_i \partial^\mu c_i + \frac{c_i}{4
  \pi} \int F^a \wedge F^a.
\ee
For simplicity we have not included mixing terms: these will not
greatly affect the discussion.

The simplest toy model is that of a factorisable toroidal orientifold, with
K\"ahler potential
\bea
\mc{K} & =  & - \ln \left( (T_1 + \bar{T}_1)(T_2 + \bar{T}_2)(T_3 + \bar{T}_3) \right) \nonumber \\
& = & - \ln (T_1 + \bar{T}_1) - \ln (T_2 + \bar{T}_2) - \ln (T_3 + \bar{T}_3).
\eea
and K\"ahler metric
\be
\mc{K}_{i \bar{j}} = \left( \begin{array}{ccc} (T_1 + \bar{T}_1)^{-2} & 0 & 0 \\
0 & (T_2 + \bar{T}_2)^{-2} & 0 \\
0 & 0 & (T_3 + \bar{T}_3)^{-2} \end{array} \right).
\ee
If we denote the axions by $c_1$, $c_2$ and $c_3$, the axion kinetic terms are
\be
\frac{1}{4 \tau_1^2} \partial_\mu c_1 \partial^\mu c_1 + \frac{1}{4 \tau_2^2} \partial_\mu c_2
\partial^\mu c_2 + \frac{1}{4 \tau_3^2} \partial_\mu c_3 \partial^\mu c_3.
\ee
For definiteness, let us assume QCD is realised on cycle 1. There is no inter-axion mixing and the relevant axion Lagrangian is
\be
\frac{1}{4 \tau_1^2} \partial_\mu c_1 \partial^\mu c_1 + \frac{c_1}{4 \pi} \int F^a \wedge F^a.
\ee
If we canonically normalise $c_1' = \frac{c_1}{\sqrt{2} \tau_1}$, this becomes
\be
\frac{1}{2} \partial_\mu c_1' \partial^\mu c_1' + \frac{\sqrt{2}\tau_1}{4 \pi} c_1' \int F^a \wedge F^a.
\ee
In units where $M_P = 1$, the axion decay constant is
$$
f_a = \frac{1}{4 \pi \tau_1 \sqrt{2}} .
$$
If QCD is to be realised on this cycle, we need $\tau_1 \sim 12$, and thus $f_a \sim 10^{16} \textrm{GeV}$.
Going beyond this toy example, we recall that in general the K\"ahler metric was given by
(\ref{GeneralKahlerP}),
\be
\mc{K}_{i \bar{j}} = \frac{G_{i \bar{j}}^{-1}}{\mc{V}^2}, \qquad \qquad
G_{i \bar{j}} = -\frac{3}{2} \left( \frac{k_{ijk}t^k}{\mc{V}} - \frac{3}{2} \frac{k_{imn}t^m t^n k_{jpq} t^p t^q}{\mc{V}^2} \right).
\ee
If all cycles are string scale in magnitude, then $\mc{K}_{i \bar{j}}
\sim \mc{O}(1)$ and it is impossible to lower the axion
decay constant substantially through canonical normalisation. The same conclusion
applies: $f_a \gtrsim 10^{16} \textrm{GeV}$. This conclusion
is unsuprising: the axionic coupling to matter is a stringy
coupling, and so we expect $f_a$ to be comparable to the string scale.
If the string and Planck scales are identical, $f_a$ cannot lie within
the allowed window.

If we lower the string scale, phenomenological values for $f_a$ can be
achieved. To analyse this, let us return to the toy model of
(\ref{VolExpression}).
We recall the K\"ahler potential was
\be
\mc{K} =  -2 \ln \left( (T_1 + \bar{T_1})^{\frac{3}{2}} - (T_2 + \bar{T}_2)^{\frac{3}{2}} - (T_3 + \bar{T}_3)^{\frac{3}{2}} \right).
\ee
The K\"ahler metric for this model is
\be
\label{KMetricfa}
\mc{K}_{i \bar{j}} = \left( \begin{array}{ccc}
\vspace{0.2cm}
\frac{-3}{2 \sqrt{2 \tau_1}\mc{V}} + \frac{9 \tau_1}{\mc{V}^2} &
-\frac{9 \sqrt{\tau_2}}{2 \mc{V}^{5/3}} & -\frac{9 \sqrt{\tau_3}}{2
  \mc{V}^{5/3}} \\
\vspace{0.2cm}
-\frac{9 \sqrt{\tau_2}}{2 \mc{V}^{5/3}} &
\frac{3}{2 \sqrt{2 \tau_2} \mc{V}} + \frac{9 \tau_2}{\mc{V}^2} & \frac{9 \sqrt{\tau_2
    \tau_3}}{\mc{V}^2} \\
-\frac{9 \sqrt{\tau_3}}{2 \mc{V}^{5/3}} & \frac{9 \sqrt{\tau_2 \tau_3}}{\mc{V}^2}
& \frac{3}{2 \sqrt{2 \tau_3} \mc{V}} + \frac{9 \tau_3}{\mc{V}^2} \end{array} \right).
\ee
The axion kinetic terms are $\mc{K}_{i \bar{j}} \partial_\mu c_i
\partial^\mu c_j$. At small volumes there is
substantial mixing between the axions $c_1$, $c_2$ and $c_3$. However,
in the limit $\mc{V} \to \infty$ with $\tau_1 \gg \tau_2, \tau_3$,
the K\"ahler metric takes the schematic form
\be
\mc{K}_{i \bar{j}} \sim \left( \begin{array}{ccc} \mc{V}^{-4/3} &
\mc{V}^{-5/3} & \mc{V}^{-5/3} \\
\mc{V}^{-5/3} & \mc{V}^{-1} & \mc{V}^{-2} \\
\mc{V}^{-5/3} & \mc{V}^{-2}
& \mc{V}^{-1} \end{array} \right),
\ee
and is to a good approximation diagonal.
The requirement $\tau_1 \gg 1$ implies that QCD cannot be realised
on branes wrapping cycle 1, as the resulting field theory is far too
weakly coupled.
However, if $\tau_2 \sim \tau_3 \sim 10$ we may realise QCD by wrapping branes on one of these
cycles.
The resulting axion decay constant is
\be
f_a \sim \frac{\sqrt{\mc{K}_{3 \bar{3}}}}{4 \pi} M_P \sim \frac{\mc{O}(1)}{4 \pi \sqrt{\mc{V}}} M_P.
\ee
Thus if $\mc{V} \sim 10^{14}$ and $\tau_3 \sim 10$,
the QCD gauge coupling is correct and the axion decay constant $f_a \sim 10^{10} \textrm{GeV}$
lies within the narrow phenomenological window. Up to $\mc{O}(1)$
factors, the string and Planck scales are related by
\be
m_s = \frac{g_s M_P}{\sqrt{\mc{V}}}.
\ee
Such a large volume corresponds to lowering the string scale to $m_s
\sim 10^{11} \textrm{GeV}$.
The lowered axion decay constant is easy to understand physically.
$f_a$ measures the axion-matter coupling, which is an effect localised
around the small QCD cycle.
Thus the only scale it is sensitive to is the string scale, and so up
to numerical factors $f_a \sim m_s$.

The above is a very particular limit of moduli space, with one cycle
taken extremely large while all others are
only marginally larger than the string scale. It would thus be essentially a curiosity if it were not also
the exact regime in which the moduli are stabilised in the compactifications of
\cite{hepth0502058, hepth0505076} reviewed above. As the stabilised volume is exponentially sensitive to the stabilised dilaton,
\emph{a priori} the string scale can lie anywhere between the Planck and TeV scales. There is no difficulty,
and no fine-tuning, in stabilising the volume so as to achieve an intermediate string scale.

The above result on $f_a$ is independent of whether an axion remains
massless or not. As described in section \ref{KKLTAxionSection}, the simplest version of the scenarios of
\cite{hepth0502058, hepth0505076} makes all axions far too heavy
to solve the strong CP problem.
In sections \ref{secnonsusyminima} and \ref{sechigherinstanton} we have described the necessary modifications to this
scenario such that a massless QCD axion will survive to solve the
strong CP problem. Combining this with
the above, we have for the first time given a procedure to stabilise
all moduli while ensuring a
QCD axion exists with a phenomenologically allowed value for
$f_a$.

In itself this is interesting, as axions within the phenomenological
window have always been hard to achieve in string compactifications.
However, this scenario compels a further very interesting relationship between the axion decay constant
and the (visible) supersymmetry breaking scale.

\subsection{Relation to Supersymmetry Breaking Scale}

We argued earlier that if a QCD axion is to be present in IIB flux
compactifications, QCD must be realised on a stack of D7-branes.
We have also described how to
stabilise the moduli such that a QCD axion can exist with a
phenomenologically allowed decay constant.

By definition,
any scenario of moduli stabilisation determines the
moduli vevs and masses.
However,
in general much more information can be extracted.
If the moduli potential breaks supersymmetry, it will generate
soft supersymmetry breaking terms in the visible sector.
In gravity-mediated scenarios, it is these that dominate the soft MSSM
Lagrangian.
It is an old and important problem to go from the moduli potential to
the soft terms.
The analysis of this problem was initiated in the heterotic string.
As full moduli stabilisation in that context is hard, the relevant
moduli potential was unknown.
Progress
was nonetheless achieved by parametrising supersymmetry breaking
as S,T or U-dominated, depending on the moduli multiplet (dilaton, K\"ahler or complex structure) in which the
dominant F-term occurred. This allowed a classification of supersymmetry breaking possibilities even in the absence
of a full moduli potential.

This question has now been revisited in IIB flux compactifications,
where the moduli potential is under better control and
full stabilisation is achievable (at least at the level of effective field theory).
Instead of having to assume the structure of the F-terms, they may now be computed directly from the moduli
potential. However, the problem is still subtle as the vacuum is
originally AdS and the form of the (necessary) uplift to Minkowksi space can
introduce extra contributions to the F-terms.
In the KKLT scenario, the AdS minimum is supersymmetric and all details of the
supersymmetry breaking therefore lie in the uplift. Unfortunately this
is the part of the construction under least control.
Using some particular assumptions about the uplift potential,
(gravity-mediated) supersymmetry breaking in this scenario has been
studied in \cite{hepth0503216}.

For the exponentially large volume compactifications,
the structure of supersymmetry breaking has been studied in
\cite{hepth0505076, hepph0512081}. The original AdS minimum is
non-supersymmetric,
and the structure of the F-terms is essentially
inherited from the no-scale potential of \cite{gkp}. While the uplift
is again not well controlled,
fortunately its effects
are subdominant.
This arises because
the magnitude of the vacuum energy in the AdS minimum is $\mc{O}\left(\frac{1}{\mc{V}^3}\right)$, whereas from the
F-terms it `ought' to be $\mc{O}\left(\frac{1}{\mc{V}^2}\right)$, the
difference of course being accounted for by the no-scale cancellation. This implies that `extra' F-terms required to
lift the minimum to Minkowski space will be hierarchically smaller
than those already present in the AdS minimum.

No-scale breaking corresponds to K\"ahler-domination,
with the dominant F-term associated with the multiplet controlling the
overall volume \cite{hepth0505076}.
The structure of the soft terms on D7-branes for the no-scale models
of \cite{gkp} has been well-studied.
Supersymmetry breaking is transmitted to the observable sector through gravitational interactions, and
the 3-form fluxes present induce soft terms on the worldvolume theory
of the D7 branes\cite{hepth0406092, hepth0408036}. Here we are only
concerned with the results rather than the calculational details:
the scalar and gaugino masses are found to be
\be
m_{D7} \sim M_{D7} \sim m_{3/2} = e^{\mc{K}/2} W \sim \frac{M_P}{\mc{V}}.
\ee
If the internal volume $\mc{V}$ is exponentially large, it is the prime
determinant of the scale of the soft terms. Other, more
model-dependent, factors will also be present, but these will only be
$\mc{O}(1)$ effects relevant for the detailed structure but not the
overall scale.
As the volume also sets the scale of the axion decay constant, this
implies a numerical relationship between these two quantities. From the fact that
$$
f_a \sim \frac{M_P}{\sqrt{\mc{V}}} \qquad \textrm{ and } \qquad m_{soft} \sim \frac{M_P}{\mc{V}},
$$
it follows that up to numerical factors
\be
f_a = \sqrt{ M_P m_{soft}}.
\ee
Thus in such models the axion decay constant is compelled to be
the geometric mean of the Planck scale and the (visible) supersymmetry breaking scale.
This is a striking result, as the two pieces of physics are \emph{a
  priori} entirely unrelated. The origin of this can be traced to the intermediate
string scale, which has been argued to have interesting
phenomenological properties\cite{hepph9810535}.

\section{Conclusions}

The purpose of this paper has been to investigate the conditions under
which a QCD axion, ideally with a phenomenological value for $f_a$,
may coexist with stabilised moduli in string compactifications. This
divides into two questions: first, how to stabilise the moduli such
that a massless axion survives, and secondly, how to obtain allowed
values of $f_a$, $10^9 \textrm{GeV} < f_a < 10^{12} \textrm{GeV}$.

In the context of the first question, we have shown that the simplest
version of many moduli stabilisation scenarios do not
contain any light axions.
We also have a negative result, in that
supersymmetric moduli stabilisation is disfavoured: there exist no
supersymmetric minima of the F-term potential with flat axionic
directions.
Even if AdS stability is present due to the Breitenlohner-Freedman
bound, the tachyons must be removed by the time we are in Minkowski
space. Performing this step requires a much greater technical understanding of uplifting
AdS vacua to Minkowksi space than is currently available, and so it is
unclear how relevant the original supersymmetric AdS solution is.

This result is pure $\mc{N}=1$ supergravity and so makes no assumptions
about the particular string model considered. It thus applies to all
string compactifications, and in particular shows that in many of the
supersymmetric IIA flux compactifications considered recently the complex structure moduli sector
is heavily tachyonic.

There is no no-go theorem on nonsupersymmetric minima with massless axions,
and so these may be preferred.
In the context of the large volume compactifications of
\cite{hepth0502058, hepth0505076}, we outlined
how to stabilise moduli while keeping axions massless.
Here we had to rely on K\"ahler corrections that will become
important as a cycle collapses to zero size. While unfortunately not
much is known about these, our main requirement was simply that they
exist.
Clearly progress in determining the form of such corrections
would be very interesting. We also specified the extent to which
subleading higher-order instantons must be absent in order for a leading-order
axion to solve the strong CP problem.

We note this result also favours gravity-mediated supersymmetry breaking. If
the moduli potential must break supersymmetry in order to solve the
strong CP problem, then this suggests that supersymmetry should be broken at
the string scale. Gravity mediation therefore always contributes to the visible
soft terms and, unless the string scale is lowered to $\mc{O}(1000
\textrm{TeV})$, will dominate over gauge mediated effects. An
intermediate string scale may then be preferred in order to obtain
TeV-scale soft terms.

In the context of the second question, the fact that $f_a$ is
hierarchically lower than the Planck scale implies that
compactifications with $m_s \sim M_P$
are unlikely to give allowed values for $f_a$.
In the compactifications of \cite{hepth0502058, hepth0505076}
the string scale is hierarchically lower than the Planck
scale. In these models, $f_a \sim M_s$ and $M_{SUSY} \sim
\frac{M_s^2}{M_P}$.
An intermediate string scale therefore gives both
$10^9 \textrm{GeV} < f_a < 10^{12} \textrm{GeV}$ and visible susy
breaking at $\mc{O}(1 \textrm{TeV})$.
It is hard to find models with
phenomenological values for $f_a$, and so it is very
interesting that in the above model this also implies TeV-scale
supersymmetry breaking. We also emphasise that as very large volumes
arise naturally in this model there is no fine-tuning problem in
having $m_{SUSY} \sim \mc{O}(1 \textrm{TeV})$.

Moduli stabilisation and
the landscape have received much discussion recently. Technically, the landscape large numbers of $\mc{O}(10^{500})$
arise from the very many ways of stabilising moduli.
It is clearly necessary to find a handle for dealing with such
numbers. One developed approach is statistical (\cite{hepth0303194}
and references thereto). Without explicitly constructing examples of
vacua, this aims at framing and answering questions about what is and what is not
possible. However, it may be very difficult to identify the right vacuum:
it has been recently argued that the problem of finding a vacuum with the right
cosmological constant is NP hard \cite{hepth0602072}.

A more general point argued here is that in the context of the
landscape the strong CP problem may serve as
an \emph{experimentum crucis}. Assuming that the solution to the strong CP
problem is a Peccei-Quinn axion and that string theory is a correct
description of nature, this is a solution that is
extremely sensitive to the physics of moduli stabilisation.
Requiring an axion to remain (essentially) massless while all
other moduli are stabilised is a technically clean problem directly
addressing the issue of vacuum selection.
Indeed, as seen above imposing this requirement directly rules
out many scenarios of moduli stabilisation.
The further condition $10^9 \textrm{GeV} < f_a < 10^{12}
\textrm{GeV}$ is even more constraining: we have described one
approach to this above and it would be very interesting to analyse others.

\section*{Acknowledgements}

I am funded by EPSRC and Trinity College, Cambridge. For conversations
related to this work, I would like to thank M. Berg, C. Burgess, M. Dine, S. Kachru,
F. Quevedo and P. Svrcek.
I am also particularly grateful to S. Kachru and F.Quevedo for commenting on drafts
of this manuscript. This work was initiated in the stimulating atmosphere
of the Stanford Institute of Theoretical Physics, whose hospitality I gratefully acknowledge.

\begin{appendix}

\section{Constraining Corrections to the K\"ahler Potential}

Here I discuss how to constrain the form of corrections to the
K\"ahler potential. The tree-level K\"ahler
potential used for our toy model was
\be
\mc{K} = - 2 \ln \left((T_1 + \bar{T}_1)^{3/2} - (T_2 + \bar{T}_2)^{3/2} -
(T_3 + \bar{T}_3)^{3/2}\right).
\ee
This geometry has one overall K\"ahler mode ($T_1$) and two blow-ups
($T_2$ and $T_3$). We specialise to our limit of interest
$\tau_1 \gg \tau_2, \tau_3 > 1$, with $\tau_i = \textrm{Re}(T_i)$, where
the resulting K\"ahler metric has the form (neglecting terms
subleading in $\mc{V}$)
\be
\label{KMetric}
\mc{K}_{i \bar{j}} = \left( \begin{array}{ccc}
\vspace{0.2cm}
\frac{3}{\mc{V}^{4/3}} &
-\frac{9 \sqrt{\tau_2}}{2 \mc{V}^{5/3}} & -\frac{9 \sqrt{\tau_3}}{2
  \mc{V}^{5/3}} \\
\vspace{0.2cm}
-\frac{9 \sqrt{\tau_2}}{2 \mc{V}^{5/3}} &
\frac{3}{2 \sqrt{2 \tau_2} \mc{V}}& \frac{9 \sqrt{\tau_2
    \tau_3}}{\mc{V}^2} \\
-\frac{9 \sqrt{\tau_3}}{2 \mc{V}^{5/3}} & \frac{9 \sqrt{\tau_2 \tau_3}}{\mc{V}^2}
& \frac{3}{2 \sqrt{2 \tau_3} \mc{V}} \end{array} \right).
\ee
Any correction to the K\"ahler potential will also generate corrections
to the K\"ahler metric (\ref{KMetric}). As such corrections are
perturbative, they may arise either from
$\alpha'$ effects or loop effects. On physical grounds, we expect that
such corrections will be subdominant to the tree-level metric in the
regime - large overall volume, large cycle volumes and weak coupling -
where both worldsheet and quantum corrections ought to be least important.

We can apply this condition to restrict the form of potential
corrections to $\mc{K}$.
For example, consider the possible correction
\be
\mc{K} + \delta \mc{K} = - 2 \ln (\mc{V}) +
\frac{\epsilon \sqrt{2 \tau_2}}{\mc{V}^{\alpha}},
\ee
with $0 < \alpha < 1$. The $2 \bar{2}$ component of the corrected K\"ahler metric is
\be
\mc{K}_{2 \bar{2}} + \delta \mc{K}_{2 \bar{2}} = \frac{3}{2 \sqrt{2 \tau_2} \mc{V}} -
\frac{\epsilon}{8 \sqrt{2} \tau_2^{3/2} \mc{V}^{\alpha}}.
\ee
As $\alpha < 1$, the correction to the kinetic term
would always dominate the
tree-level term in the limit $\mc{V} \gg 1$. This seems implausible as a large-volume limit ought
to make the correction less, rather than more, important.

A similar comment applies to a correction
\be
\mc{K} + \delta \mc{K} = - 2 \ln (\mc{V}) +
\frac{\epsilon \tau_2^2}{\mc{V}},
\ee
which leads to
\be
\mc{K}_{2 \bar{2}} + \delta \mc{K}_{2 \bar{2}} = \frac{3}{2\sqrt{2 \tau_2} \mc{V}} +
\frac{\epsilon}{4 \mc{V}}.
\ee
In this case, as we take $\tau_2$ large, the correction becomes
increasingly dominant over the tree-level term.
Given that large $\tau_2$ reduces both the curvature of this cycle and
the gauge coupling on any brane wrapping it,
we would again expect exactly opposite behaviour to occur.

Finally, we could also consider the correction
\be
\mc{K} + \delta \mc{K} = - 2 \ln (\mc{V}) +
\frac{2 \epsilon \tau_2}{\mc{V}^{\alpha}},
\ee
with $0 < \alpha < 1$.
In this case $\delta K_{2 \bar{2}}$ is subleading to $\mc{K}_{2
  \bar{2}}$ in the classical limit. However, if we consider the $1
\bar{2}$ component we now have
\be
\mc{K}_{1 \bar{2}} + \delta \mc{K}_{1 \bar{2}} =
-\frac{9 \sqrt{\tau_2}}{2 \mc{V}^{5/3}} + \frac{3 \epsilon}{2
  \mc{V}^{\alpha + 2/3}},
\ee
and the correction again dominates in the large-volume limit.

The correction used in the body of the paper,
\be
\label{KCorrectionUsed}
\mc{K} + \delta \mc{K} = - 2 \ln (\mc{V}) +
\frac{\epsilon \sqrt{\tau_2}}{\mc{V}} + \frac{\epsilon \sqrt{\tau_3}}{\mc{V}},
\ee
does not suffer from the above problems. As a correction to the K\"ahler
metric, it gives
\bea
\mc{K}_{2 \bar{2}} + \delta \mc{K}_{2 \bar{2}} & = & \frac{3}{2
  \sqrt{2 \tau_2} \mc{V}} -
\frac{\epsilon}{16 \tau_2^{3/2} \mc{V}}, \\
\mc{K}_{3 \bar{3}} + \delta \mc{K}_{3 \bar{3}} & = & \frac{3}{2
  \sqrt{2 \tau_3} \mc{V}} -
\frac{\epsilon}{16 \tau_3^{3/2} \mc{V}}.
\eea
Such corrections are
suppressed compared to the tree-level term by a factor $\tau^{-1}$, i.e. $g^2$ of the
field theory on the brane.
Unlike those considered above, these corrections are well-behaved
(i.e. subdominant) in the classical limit, and there
does not exist a `bad' scaling limit.

We note that for the case where K\"ahler corrections have been computed
\cite{hepth0508043}, the correction does fall into this form.
We only focus on the K\"ahler moduli dependence: the full expressions
can be found in \cite{hepth0508043}.
The correction gives
\be
\mc{K} + \delta \mc{K} = - \ln (T_1 + \bar{T}_1) - \ln (T_2 +
\bar{T}_2) - \ln (T_2 + \bar{T}_3) + \sum_{i=1}^3
\frac{\epsilon_i}{T_i + \bar{T}_i},
\ee
with for example
\be
\mc{K}_{1 \bar{1}} + \delta \mc{K}_{1 \bar{1}} = \frac{1}{4 \tau_1^2} +
\frac{\epsilon_1}{4 \tau_1^3}.
\ee
The loop-corrected K\"ahler metric is suppressed by a factor
$\tau_1^{-1} = g^2$.

The point of this discussion is that
the form of possible corrections to the K\"ahler potential can be
heavily constrained by the reasonable requirement that in a classical (large
volume, weak coupling) limit, corrections to the metric become increasingly
subdominant to tree-level terms: simply because K\"ahler
corrections are very hard to calculate does not make us entirely
ignorant of their form.
In particular, denoting the `small', blow-up moduli by $\tau_i$, these considerations exclude
corrections of the form
\be
\mc{K} + \delta K = - 2 \ln (\mc{V}) +
\frac{f(\tau_i)}{\mc{V}^{\alpha}},
\ee
with $\alpha < 1$, as in a classical, large volume limit there will be metric
components whose correction dominates the tree-level term.

This motivates our use of the correction (\ref{KCorrectionUsed}): it
comes from a reasonable assumption about the form of the corrections
to the K\"ahler metric, and is consistently subdominant in the
classical limit.

\end{appendix}

\end{document}